\documentstyle[aps,epsfig,twocolumn]{revtex}
\begin{document}
\draft
\flushbottom
\twocolumn[
\hsize\textwidth\columnwidth\hsize\csname @twocolumnfalse\endcsname
\title{
Electron-electron interactions in graphene sheets}
\author{J. Gonz\'alez$^{\dag}$, F. Guinea$^{\ddag}$ and
M. A. H. Vozmediano$^*$ \\}
\address{
$^{\dag}$ 
Instituto de Estructura de la Materia. 
Consejo Superior de Investigaciones Cient{\'\i}ficas. 
Serrano 123, 28006 Madrid. Spain. \\
$^{\ddag}$
Instituto de Ciencia de Materiales.
Consejo Superior de Investigaciones Cient{\'\i}ficas.
Cantoblanco. 28049 Madrid. Spain.  \\
$^*$
Departamento de Matem\'aticas. Universidad Carlos III.
Butarque 15. Legan\'es. 28913 Madrid. Spain.}
\date{\today}
\maketitle
\tightenlines
\widetext
\advance\leftskip by 57pt
\advance\rightskip by 57pt

\begin{abstract}
The effects of the electron-electron interactions in a graphene layer
are investigated. It is shown that short range couplings are
irrelevant, and scale towards
zero at low energies, due to the vanishing of density of states
at the Fermi level.
Topological disorder enhances the density of states, and can lead
to instabilities. In the presence of sufficiently strong repulsive
interactions, p-wave superconductivity can emerge.  
 
\end{abstract}

\pacs{75.10.Jm, 75.10.Lp, 75.30.Ds.}
]
\narrowtext
\tightenlines
\section{Introduction.}
Recent experiments\cite{Ketal00,Ketal00b} report the existence
of ferromagnetic and superconducting fluctuations in graphite at
unexpectedly high temperatures ($T \sim 100 - 300$K). The coexistence
of both types of fluctuations suggests a common electronic
origin for them.

Motivated by these observations, we present here a study of the possible
electronic instabilities of a single graphene sheet. Isolated graphene
has the convenient property that the electronic states near the
Fermi level can be described in simple terms. By symmetry, the lower and
upper bands touch at the corners of the hexagonal Brillouin zone.
Near these points, the dispersion relation is isotropic and linear,
$\epsilon_{\bf \vec{k}} = v_F | {\bf \vec{k}} |$, where $v_F$ 
is the Fermi velocity.
The density of states at the Fermi level is strictly zero, and it rises
linearly in energy.
An effective long wavelength description of these electronic
states can be written in terms of the Dirac equation in two dimensions
(see below).  

The fact that a single graphene sheet is a semimetal modifies
significantly the screening of the Coulomb interaction\cite{GGV94}.
An effective low energy hamiltonian can be written, which can be treated
by Renormalization Group methods\cite{S94,P92}. It can be shown 
rigourously that the Coulomb interaction is a marginal interaction,
which scales to zero at low energies or long wavelengths. 
At intermediate scales, however, the quasiparticle lifetime 
does not follow the usual $\epsilon^2$ dependence of Landau's
theory of a Fermi liquid, but scales as $| \epsilon |$\cite{GGV96},
in agreement with experiments\cite{Yetal96}.
The RG approach is, in principle, valid in the weak coupling regime,
($e^2 / ( \epsilon_0 v_F ) \ll 1 $, where $e$ is the electric charge
and $\epsilon_0$ is the dielectric constant. By using a RPA summation
of diagrams, it can be shown that the low energy properties 
are not changed throughout the entire range of couplings\cite{GGV99}.
 
The previous work mentioned earlier analyzed the  
the small momentum scattering 
due to the long range Coulomb interaction, as it is the only one which leads to
logarithmically divergent perturbative corrections. Some electronic
instabilities, like anisotropic superconductivity, requires
the existence of short range interactions with significant
strength at finite wavevectors.  We analyze in this work the role
of these interactions in inducing instabilities of the 
electronic system. The next section describes the model.
Then, the Renormalization Group equations for the different
iteractions are written. In section IV, the role of topological
disorder is analyzed, as it can lead to changes in the density
of states which modify the scaling equations 
obtained earlier. The main conclusions are presented
in section V.
\section{The model.}
\subsection{Intralayer couplings.}
\begin{figure}
\centerline{\epsfig{file=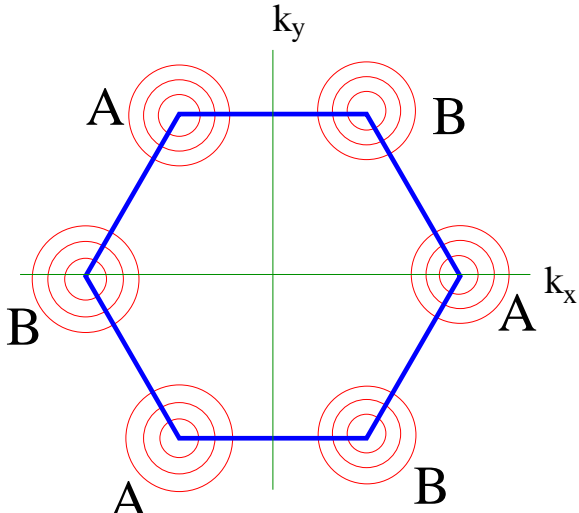,width=3in}}
\caption{Sketch of the Brillouin Zone of a graphene sheet, and the
band dispersion near the Fermi energy.}
\label{fig1}
\end{figure}

We analyze the low energy properties of a graphene sheet.
We will only consider the modifications due to interactions
and disorder in the low energy properties of the system. 
Thus, we need to describe the low energy electronic states.
A graphene sheet has an hexagonal symmetry with two atoms
per unit cell. The carbon atoms have four valence orbitals.
Three of them build the sp$^2$ bonds which give
rigidity to the structure. The third orbital gives rise to 
the valence and conduction bands. These bands touch at the 
two inequivalent corners
of the Brillouin zone (see Fig.[\ref{fig1}]). 
From symmetry considerations, these
bands are isotropic, and depend linearly on the wavevector.

It can be shown that, in the long wavelength limit, the
electronic wavefunctions near the corners of the Brillouin
Zone are well described in terms of the two dimensional 
Dirac equation. Each of the two inequivalent points 
requires two Dirac spinors, each of them with its spin index. 
In the long wavelength limit, the
The Fermi velocity, $v_F$, can be expressed in terms of the
matrix elements between nearest neighbor $\pi$ orbitals, $t$,
as $v_F = (3 t a)/2 $, where $a$ is the C-C distance. 

Because of the collapse of the Fermi surface to isolated points,
the kinematics are much simpler than the corresponding analysis
for two ^^ ^^ hot spots " in a square lattice\cite{GGV00}.
\twocolumn
[\hsize\textwidth\columnwidth\hsize\csname@twocolumnfalse%
\endcsname
The hamiltonian is:
\begin{eqnarray}
{\cal H} &= &\sum_{i,s} \hbar v_F \int d^2 r \bar{\Psi}_{i,s} ( \vec{r} )
( i \sigma_x \partial_x + i \sigma_y \partial_y )
\Psi_{i,s} ( \vec{r} ) + \nonumber \\ &+ &\sum_{i,i';s,s'}
\frac{e^2}{2 \epsilon_0}
\int d^2 r_1 \int d^2 r_2 \frac{\bar{\Psi}_{i,s} ( \vec{r}_1 )
\Psi_{i,s} ( \vec{r}_1 ) \bar{\Psi}_{i',s'} ( \vec{r}_2 ) \Psi_{i',s'}
( \vec{r}_2 )}
{| \vec{r}_1 - \vec{r}_2 |} + \nonumber \\
&+ &\sum_{ s,s'; i,i'}  g_{i,s;i',s'}
\int d^2 r \bar{\Psi}_{i,s} ( \vec{r} )  
\Psi_{i,s} ( \vec{r} ) \bar{\Psi}_{i',s'} ( \vec{r} )
\Psi_{i',s'} ( \vec{r} ) + \nonumber \\
&+ &\sum_{ s,s'; i,i'}  \bar{g}_{i,s;i',s'}
\int d^2 r \bar{\Psi}_{i,s} ( \vec{r} )  
{\bf \vec{\sigma}} \Psi_{i,s} ( \vec{r} ) \bar{\Psi}_{i',s'} ( \vec{r} )
{\bf \vec{\sigma}} \Psi_{i',s'} ( \vec{r} )
\label{hamil}
\end{eqnarray}
]
where $\sigma_x$
and $\sigma_y$ are $2 \times 2$ Pauli matrices.
We have separated the long wavelength part of the Coulomb interaction from
other possible short range interactions. 

\begin{figure}
\centerline{\epsfig{file=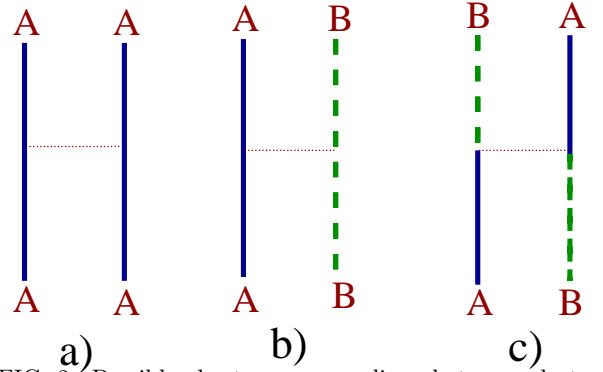,width=3in}}
\caption{Possible short range couplings between electrons near
the two Fermi points in a graphene layer. The full and
broken lines correspond to electrons in the vicinity
of each of the two Fermi points.
a) Intra-singularity
scattering ($g_{\rm intra}$). b) Inter-singularity scattering
($g_{\rm inter}$). c) Exchange scattering ($g_{\rm exchange}$).}
\label{fig2}
\end{figure}

The couplings, $g_{i,i';s,s'}$ can be classified in an
analogous way as in one dimension. The possible scattering processes
are shown in Fig.[\ref{fig2}].

Because of the linear dispersion of
the electronic states, we can use $v_F$ to transform time
scales into length scales. Then, we can express the dimensions of
all physical quantities in terms of lengths.
Within this convention, we find that the dimension  of
the electronic fields is $[ \Psi ] = l^{-1}$,
where $l$ defines a length. 
A naive power counting analysis shows that the Coulomb potential
defines a dimensionless, marginal coupling, while the $g$'s
scale as $ l $, and are irrelevant at low energies.
This effect can be traced back to the vanishing density of states
at the Fermi level.
When a single Hubbard
intrasite repulsion $U$ is considered, all interactions between
electrons of opposite spin in eq. \{\ref{hamil}\} are equal
to $U \Omega$, where $\Omega$ is the area of the unit cell, and
the interactions between electrons of parallel spin are zero.
\subsection{Interlayer couplings.}
\begin{figure}
\centerline{\epsfig{file=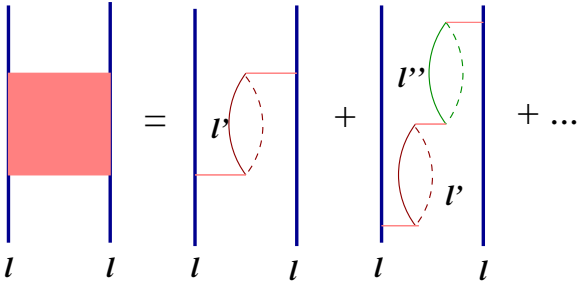,width=3in}}
\caption{Diagrammatic representation  of the Random Phase Approximation
applied to the interlayer Coulomb interaction.
The diagrams show the screening of the interaction between
two electrons in layer $l$ due to the polarization of layers
$l' , l'' ...$.}
\label{fig4}
\end{figure}

So far, we have restricted our analysis to processes within an
isolated graphene sheet. Neighboring layers are always coupled by the
Coulomb interaction. In the following, we will neglect interlayer
hopping, so as to be able to describe the electronic levels in
terms of the Dirac equation, but we include the effects of the
long range Coulomb interactions between layers.
The interlayer couplings give rise to the screening of the bare
intralayer electron-electron interaction. We will treat these 
effects within the RPA, as depicted in Fig.[\ref{fig4}], following
the analysis in\cite{GGV96}. 
\twocolumn
[\hsize\textwidth\columnwidth\hsize\csname@twocolumnfalse%
\endcsname
The intralayer interaction
becomes:
\begin{equation}
v_{scr} ( \omega , {\bf \vec{q}} ) = 
\frac{2 \pi e^2}{ \epsilon_0 | {\bf \vec{q}} | } 
\frac{\sinh ( | {\bf \vec{q}} | d )}
{\sqrt{ \left[ \cosh ( | {\bf \vec{q}} | d ) + \frac{2 \pi e^2}
{ \epsilon_0 | {\bf \vec{q}} |}
\sinh ( | {\bf \vec{q}} | d )  
\chi_0 ( \omega , {\bf \vec{q}} ) \right]^2 - 1 }}
\label{suscrpa}
\end{equation}
]
where $d$ is the distance between layers, and $\chi_0$ is the electron 
susceptibility of a single layer, given by:
\begin{equation}
\chi_0 ( \omega , {\bf \vec{q}} ) = \frac{{\bf \vec{q}}^2}{32 \pi \sqrt{
v_F^2 {\bf \vec{q}}^2 - \omega^2}}
\label{susc}
\end{equation}

The interlayer interactions are only effective 
when $| {\bf \vec{q} } | d \ll 1$.
Hence, if the lattice constant $a$
is such that $a \ll d$, they do not affect significantly the couplings between
electronic states in different Fermi points.
\section{Scaling analysis.}
\begin{figure}
\centerline{\epsfig{file=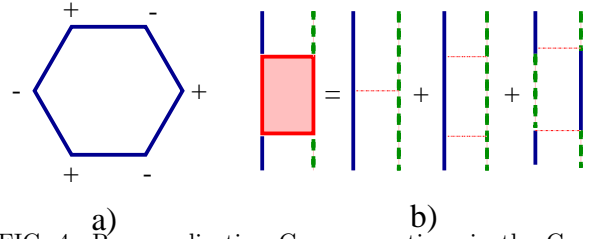,width=3in}}
\caption{Renormalization Group equations in the Cooper channel with
p-wave symmetry. a) Sketch of the order parameter in the Brillouin Zone.
b) Diagrams involved in the calculation.}
\label{fig3}
\end{figure}

In\cite{GGV99} it was shown that the electrostatic coupling,
defined as $e^2 / ( \epsilon_0 v_F )$, scales towards zero at
low energies, for all values of the interaction.
On the other hand, the existence of scattering processes
between the two inequivalent Fermi points can lead to
instabilities at intermediate couplings. Different combinations of
couplings lead to each instability. The system becomes
ferromagnetic for sufficiently large values of
$g_{\rm{intra} \perp} + g_{\rm{inter} \perp} -
g_{\rm{intra} \parallel} - g_{\rm{inter} \parallel} $, where the 
subscripts $\parallel$ and $\perp$ denote the relative
orientation between spins.
An antiferromagnetic instability is driven by
$\bar{g}_{\rm{intra} \perp} + \bar{g}_{\rm{inter} \perp} -
\bar{g}_{\rm{intra} \parallel} - \bar{g}_{\rm{inter} \parallel} $.
The superconducting phases can be s and p wave, depending on the
relative phase of the gap at the two inequivalent points.
However, for each ${\bf \vec{k}}$ near the Fermi points,
there are two electronic states, so that an additional
index can be defined in the superconducting order parameter.
Writing these two states as a two component spinor, we can write, in general: 
\begin{equation}
\Delta_{\bf \vec{k}} = \langle \Psi_{A, \uparrow , {\bf \vec{k}}}
\left( a {\cal I} + {\bf \vec{b}} \, {\bf \vec{\sigma}} \right)
\Psi_{B , \downarrow , - {\bf \vec{k}}} \rangle
\label{op}
\end{equation}
where $a$ and ${\bf \vec{b}}$ are constants. 
When the interaction is repulsive, the p-wave symmetry 
is favored ($\Delta_{\bf \vec{k}} = - \Delta_{- {\bf \vec{k}}}$),
as in a two dimensional electron system with two inequivalent
van-Hove singularities at the Fermi level\cite{GGV00}.
The corresponding coupling is $g_{{\rm inter} \perp} +
\bar{g}_{{\rm inter} \perp} - g_{{\rm exchange} \perp} -
\bar{g}_{{\rm exchange} \perp}$. The diagrams which define the flow
of these couplings are depicted in Fig.[\ref{fig3}].
The corresponding equations for the dimensionless
vertices, $\tilde{\Gamma}$'s, can be written as:
\begin{eqnarray}
\frac{\partial \tilde{\Gamma}_{\rm inter}}{\partial \log ( \Lambda )} &=
&-   d_{\tilde{\Gamma}} 
\tilde{\Gamma}_{\rm inter} - \tilde{\Gamma}_{\rm inter}^2 - 
\tilde{\Gamma}_{\rm exchange}^2 \nonumber \\
\frac{\partial \tilde{\Gamma}_{\rm exchange}}{\partial \log ( \Lambda )} &=
&-  d_{\tilde{\Gamma}} \tilde{\Gamma}_{\rm exchange} - 
2 \tilde{\Gamma}_{\rm exchange} 
\tilde{\Gamma}_{\rm inter} 
\label{flow}
\end{eqnarray}
where we are omitting spin and flavor indices for simplicity,
and $d_{\tilde{\Gamma}}$ is the (anomalous) dimension of the vortex,
which includes, among others, the effects of the wavefunction
renormalization of the fields. To lowest order, $d_{\tilde{\Gamma}} = 1$.
The first term in the r. h. s. of eqs.[\ref{flow}] is linear, and it
is absent in the flow of the couplings in the Cooper channel
in a conventional metal. It reflects the irrelevance of these
couplings in a semimetal.

The flow in this channel becomes relevant if $\tilde{\Gamma}_{\rm exchange}
\ge \tilde{\Gamma}_{\rm inter}$ and the values of the $\tilde{\Gamma}$'s are of 
order unity. Note that the cutoff is assumed to be $\Lambda \approx
v_F / a$, where $a$ is a length of the order of the lattice constant.
The dimensionful inter-Fermi points and exchange couplings induced
by the Coulomb interactions are $g_i \sim e^2 / ( \epsilon_0 a )$.
Hence, the bare vortices, 
$\tilde{\Gamma}_0 \sim e^2 / ( \epsilon_0 v_F )$. For reasonable
values of $\epsilon_0 \sim 4 - 8$, this combination is, indeed,
of order unity.
\section{Influence of disorder.}
\subsection{Topological disorder.}
\begin{figure}
\centerline{\epsfig{file=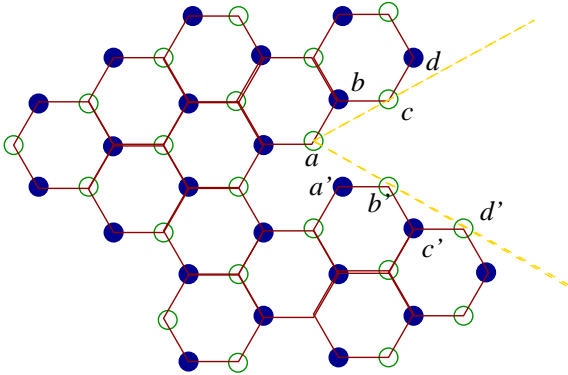,width=3in}}
\caption{Formation of a pentagonal ring in the honeycomb
lattice. Points $a, b, c, d ...$ have to be identified
with points $a', b', c', d' ...$. The defect can be seen
as a disclination, defined by the straight dashed lines.}
\label{fig5}
\end{figure}

The formation of pentagons and heptagons in the lattice, without affecting
the threefold coordination of the carbon atoms, lead to the warping of
the graphene sheets, and are responsible for the formation of curved 
fullerenes, like C$_{60}$. They can be viewed as disclinations
in the lattice, and, when circling one such defect, the 
two sublattices in the honeycomb structure are exchanged
(see Fig.[\ref{fig5}]).
The two fermion flavors defined in eq.\{\ref{hamil}\} are also exchanged
when moving around such a defect. The
scheme to incorporate this change in a continuum
description was discussed in\cite{GGV92}.
The process can be described by means
of a non Abelian gauge field, which rotates the spinors in
flavor space. The vector potential is that of a vortex at the
position of the defect, and the flux is $\pm \pi / 2$.

Dislocations can be analyzed in terms of bound disclinations,
that is, a pentagon and an heptagon located at short distances,
which define the Burgers vector of the dislocation. Thus, 
the effect of a dislocation on the electronic levels of
a graphene sheet is analogous to that of the vector potential
arising from a vortex-antivortex pair. We can extend this description\cite{G98},
and assume that a lattice distortion which rotates the lattice axis
can be parametrized by the angle of rotation, $\theta (
{\bf \vec{r}} )$,  of the local axes
with respect to a fixed reference frame. Then, this distortion
induces a gauge field such that:
\begin{equation}
 {\bf \vec{A}} ( {\bf \vec{r}} )  = 3 \nabla \theta ( {\bf \vec{r}} )
\left( \begin{array}{cc} 0 &-i \\ i &0 \end{array} \right)
\label{gauge}
\end{equation}
Thus, a random distribution of topological defects can be described 
by a (non abelian) random gauge field. The nature of the electronic
states derived from the two dimensional Dirac equation in the presence
of a  gauge field with gaussian randomness
has received a great deal of attention, as it  
also describes the effects of disorder 
in integer quantum Hall transitions\cite{Letal94}.
The disorder is defined by a single dimensionless quantity,
$\Delta$, which is proportional to the average fluctuations
of the field:
\begin{equation}
\langle {\bf \vec{A}} ( {\bf \vec{r}} ) {\bf \vec{A}} (
{\bf \vec{r}'} ) \rangle = \Delta \delta^2 ( {\bf \vec{r}} -
{\bf \vec{r}'} )
\label{delta}
\end{equation}
It is known that $\Delta$ gives rise to a marginal perturbation,
which modifies the dimensions
of the fermion fields and enhances the density of states at low
energies. A variety of analytical\cite{CMW96} and numerical 
techniques\cite{MH97} has been used to study this problem.
We will follow the Renormalization Group scheme presented
in\cite{Letal94}. 

\begin{figure}
\centerline{\epsfig{file=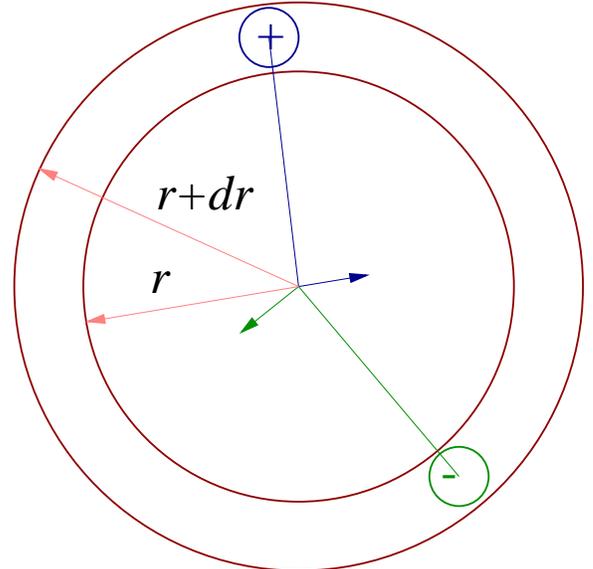,width=3in}}
\caption{Effect of vortices located at distances
between $r$ and $r + dr$ from the origin (see text for discussion).}
\label{fig6}
\end{figure}

We first analyze the statistical properties of  the gauge field
induced by topological defects. Let us assume that the graphene 
sheet is warped, and that there is a random distribution of pentagons
and heptagons, with density $n_0$ and average distance equal
to $l_0 = n_0^{-1/2}$. The fluctuations in the gauge field induced
by this distribution at a given point can be calculated by 
considering the effect of all defects located at distances
between $r$ and $r + dr$ (see Fig.[\ref{fig6}]), where $r \gg l_0$.
The number of defects of each type is $2 \pi r dr n_0$. The angle,
$\phi$, which defines their position is a random variable. The 
contribution to the x-component
of the gauge field from these vortices is:
\begin{eqnarray}
A_x^2 ( {\bf \vec{r}} = 0 )
 &= &\left( \frac{\Phi_0}{r} \right)^2 
\left[ \sum_i \cos ( \theta_i ) \right]^2
\nonumber \\
&= &\left( \frac{\Phi_0}{r} \right)^2 \frac{1}{2} \, \,  2 \pi n_0 r dr 
\end{eqnarray}
where $\Phi_0$ is the flux associated to a single vortex,
and there is a similar equation for $A_y ( {\bf \vec{r}} = 0 )$.
We now must integrate this value from $l_0$ to $R$, where $R$
is the radius of the sample. We obtain:
\begin{equation}
| {\bf \vec{A}} ( {\bf \vec{r}} = 0 ) |^2 = 
2 \pi n_0 \Phi_0^2 \log \left( \frac{R}{l_0} \right)
\end{equation}
We can assume that the vector potential at positions separated
by distances greater
than $l_0$ are not correlated. Then, from eq.\{\ref{delta}\},
we find:
\begin{equation}
\Delta = 2 \pi \Phi_0^2 \log \left( \frac{R}{l_0} \right)
\label{vortex}
\end{equation}
which diverges slowly with the size of the system. 
The previous estimate assumed that the layers had a significant amount of
curvature at distances smaller than $l_0$. We can alternatively
assume that pentagons and heptagons are bound in dislocations 
with average distance $b$. The vector field of a vortex-antivortex
dipole decays as $r^{-2}$. A similar analysis to the one leading to
eq.\{\ref{vortex}\} gives:
\begin{equation}
\Delta \propto \Phi_0^2 n_{disl} b^2
\label{disloc}
\end{equation}
where $ n_{disl}$ is the density of dislocations.

We will now assume that random fields induced by topological defects have
the same statistical properties to those with gaussian disorder with
the same value of $\Delta$, which is 
the second moment of the distribution in both cases.
Then, we can perform the Renormalization Group analysis discussed
in\cite{Letal94}. To lowest order, we find an interaction between
fermion fields in different replicas of the type:
\begin{eqnarray}
S_{int} &= &\Delta \sum_{m,n} \int 
 \left[ \bar{\Psi}_A ( {\bf
\vec{r}} , t_1 ) \Psi_B ( {\bf \vec{r}} , t_1 ) \right]_m 
\nonumber \\ &\times &\left[
\bar{\Psi}_B ( {\bf \vec{r}} , t_2 ) 
\Psi_A ( {\bf \vec{r}} , t_2 ) \right]_n  d t_1 d t_2 d {\bf \vec{r}}
\end{eqnarray}  
where $m$ and $n$ are replica indices. This interaction leads
to a logarithmically divergent self energy, which can be interpreted
as a renormalization of the density of states\cite{Letal94}.
We can include the corrections induced
by the self energy in a renormalization of the wave function,
giving rise to a change in the scaling dimension of the fields:
\begin{equation}
2 d_{\Psi} - 1 = 1 - 
\frac{\Delta}{\pi}
\label{dimension}
\end{equation}
This expression has to be inserted in eq.\{\ref{flow}\}, modifying
the flow of the couplings.

\begin{figure}
\centerline{\epsfig{file=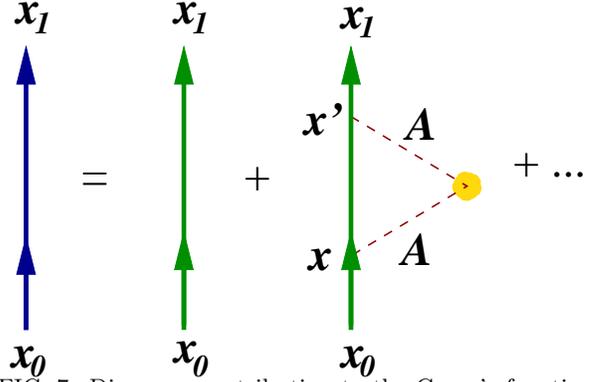,width=3in}}
\caption{Diagrams contributing to the Green's function in a disordered
electron system.}
\label{fig7}
\end{figure}

The same result can be reached by analyzing the self energy
corrections using standard techniques in the study of disordered
electrons in arbitrary dimensions\cite{E83}. To lowest order, the
first correction to the Green's function is shown in Fig.[\ref{fig7}].
This diagram leads to a self energy:
\begin{eqnarray}
\Sigma ( {\bf \vec{r}} , 
{\bf \vec{r'}} , \omega ) &\approx
&\langle G_0 ( {\bf \vec{r}} - {\bf \vec{r'}} ,
\omega ) {\bf \vec{A}} ( {\bf \vec{r}} ) {\bf \vec{A}} ( {\bf \vec{r'}} )
\rangle + \cdots  \nonumber \\
&=   &\Delta G_0 ( {\bf \vec{r}} - {\bf \vec{r'}} , \omega ) \delta^2
( {\bf \vec{r}} - {\bf \vec{r'}} ) + \cdots
\end{eqnarray}
where $G_0$ is the unperturbed Green's function. The
real part of $G_0$ behaves as $G_0 \sim \omega \log ( \Lambda /
\omega )$. Finally:
\begin{equation}
2 d_{\Psi} - 1 = 1 - \frac{\partial}{\partial \log \Lambda}
\left( \frac{\partial \Sigma}{\partial \omega} \right)
\end{equation}

The previous perturbative analysis can be generalized to arbitrary couplings,
by mapping the non interactiong fermion problem in two spatial dimensions
onto an interacting problem in 1+1 dimensions\cite{NTW94}. At energies 
below a scale $\lambda \sim \Lambda \exp [ - \pi / ( 2 \Delta ) ]$, the
backscattering between the two Fermi points leads to a scaling
dimension which is independent of the disorder, $d_{\Psi} = 1/7$.

\subsection{Substitutional and site disorder.}
It can be shown that substitutional and site disorder can be incorporated
into the Dirac equation through a change in the local chemical potential
and the appearance of a mass term\cite{CMW96}.
Disorder of these types, with a gaussian distribution, defines a marginally
relevant perturbation\cite{CMW96}. Within the perturbative RG scheme
described in the previous subsection, it can be shown that this
perturbation leads to logarithmic corrections to the site-diagonal
self energy, which can be incorporated into a renormalization of
the wavefunction. Moreover, deviations from a gaussian distribution 
are relevant perturbations, which modify the results.

Thus, while to lowest order substitutional disorder shows similar
characteristics as
topological disorder, higher order corrections lead to
significant modifications, which, in addition, depend on the
type of disorder\cite{Aetal00}. For instance, 
a sharp divergence of the density of states has been found\cite{PL98},
or a suppression of the density of states at low energies\cite{SF99}. 

In graphite, we do not expect a high concentration of charged impurities,
which will lead to a strong site disorder. Randomness in bond lengths
leads to non diagonal disorder, which can be included in the topological
disorder discussed in the previous subsection. Hence, we expect 
diagonal disorder in pure graphite to be small, leading to minor corrections
to the dependence of the density of states on energy, even when
non perturbative terms are included\cite{Aetal00}.
\section{Discussion.}
\subsection{Other effects not included in the model.}
We have neglected the effects due to phonons, although,
in principle, they can be incorporated into the framework used here.
Our main purpose is the study of instabilities towards
ground states which exhibit magnetism or anisotropic superconductivity.
We assume that the electron-phonon interaction will not
change qualitatively the  possible existence of these instabilities.

We have also not consider the interlayer hopping, $t_{\perp}$. Within the RG
scheme, coherent interlayer hopping is 
a relevant perturbation\cite{W90}, leading
to three dimensional behavior at low energies or temperatures.
On the other hand, due to the vanishing of the density of states of
a graphene layer, incoherent hopping between layers is irrelevant
(note that it is a marginal perturbation in systems with a
finite density of states\cite{KF92}). 

In the presence of coherent interlayer hopping, our analysis is
valid only at scales higher than $t_{\perp}$, which has been estimated,
by band structure calculations 
to be $t_{\perp} \approx 0.27$eV\cite{Z78}. This bare value
will be reduced by the many body effects, and the wave function renormalization
considered here. However, in a perfect system, the validity of our calculations
are limited to a range between $t \approx 2.4$eV and the
renormalized value of $t_{\perp}$.

The coherent interlayer hopping can modify our results in various ways:
i) The coupling between layers induces a crossover to 3D behavior,
enhancing the 2D instabilities discussed here. ii) The dispersion
of the electronic bands in the third dimension leads to the
existence of small electron and hole pockets, increasing the density
of states. If the couplings are not modified, this finite
density of states will also strenghten the instabilities.
ii) The density of states at the Fermi level induces metallic 
screening, and changes the interactions at low energies.
It is unclear to us how our results are modified in case iii).
 
Our calculations have a wider range of validity in the presence of
disorder, where coherent hopping over distances longer than the 
electronic mean free path is suppressed. As mentioned earlier,
incoherent local hopping can be considered an irrelevant perturbation
which should not modify qualitatively the results presented here. 
\subsection{Analysis of the couplings.}
Our analysis considers the role of electron-electron interactions 
in a graphene layer. Spin dependent interactions, like a Hubbard
on site term, naturally lead to magnetic phases. In the absence of disorder,
a minimum value for the Hubbard repulsion is required before
the onset of antiferromagnetism\cite{ST92},
in agreement with the analysis presented here. This phase, however, lacks
experimental confirmation. It is also known that,
within the Hartree-Fock approximation,  a nearest 
neighbor repulsion, $V$, induces a charge density wave ground state,
if $U - 3 V < 0$, and $U$ and $V$ are sufficiently large\cite{TH92,WL96,LCM00}.
In addition, there is a region in the phase diagram where $U$ and $V$ 
almost cancel, leading to a paramagnetic ground state. Realistic values
of these parameters sugest that a graphene layer lies in this 
region\cite{TH92,WL96,LCM00}. 
It is reasonable that longer range correlations can 
make this state unstable. These calculations do not consider longer range
interactions. For decoupled graphene layers, the vanishing of the
density of states at the Fermi level leads to the absence of
metallic screening, so that spin independent,
long range interactions are expected.
\subsection{Low temperature phases.}
We have considered the possibility of ferro- and antiferomagnetism,
and p wave superconductivity as the most likely low temperature
phases. The competition between them depends on the spin dependence
of the interactions. Spin independent couplings favor superconductivity,
while a strong spin dependence, like the on site repulsion of the
Hubbard model, will lead to a magnetic ground state. Finally, the
relative stability of ferro- and antiferromagnetism depends,
among other things, on the existence of an underlying bipartite lattice.
In the presence of a sufficiently strong topological disorder, we expect
that ferromagnetism will prevail over antiferromagnetism, as the
existence of pentagons and heptagons leads to the frustration
of antiferromagnetic order. The same argument can be applied to the
charge density wave state considered in\cite{TH92,WL96}.

A detailed study of the competition between ferromagnetism and p-wave
superconductivity lies beyond the scope of this work. 
It depends on the balance between the on site, spin dependent interactions,
and the longer range, spin independent couplings. Ferromagnetism is favored
by the existence of a sufficiently strong forward scattering between
electrons of opposite spin, at momentum transfer ${\bf \vec{q}} \approx 0$.
This coupling depends on the nature of the screening, which, in turn,
depends on the density of states near the Fermi level, and on the 
degree of disorder. On the other hand, if the main interactions
are spin independent, ferromagnetism will be suppressed, and
the leading instability is p-wave superconductivity.
\subsection{General features of the possible superconducting instability.}
In the following, we will focus discuss some qualitative features of the 
superconducting transition.
A quatitative estimate of the critical temperature is beyond the scope
of our RG scheme, although we can discuss the dependence
of $T_c$ on various quantities.
 
It is first interesting to note
that superconductivity at low temperatures was observed in 
graphite intercalation compounds\cite{Hetal65}.
The origin of this
superconductivity is not completely understood.  The 
critical field shows an anomalous 
dependence on temperature\cite{KT80}, unlike
in conventional s-wave superconductors. This dependence has been explained 
in terms of a two band model\cite{JDC91}. This model is similar to the
two point model discussed here, except that the two bands considered 
in\cite{JDC91} correspond to a carbon and a dopand band. The temperature
dependence of the critical field should be, however, similar
in the two cases. 

The critical temperature at which an instability described by
eq.\{\ref{flow}\} sets in is:
\begin{equation}
T_c = \Lambda \left( \frac{\tilde{\Gamma}_0
 - d_{\tilde{\Gamma}}}{\tilde{\Gamma}_0} \right)^\frac{1}
{d_{\tilde{\Gamma}}}
\label{Tc}
\end{equation}
where $\tilde{\Gamma}$ is the appropiate vortex required
to drive the instability. There is a transition if
$ \tilde{\Gamma}_0 > \tilde{\Gamma}_c 
= d_{\tilde{\Gamma}}$. For $d_{\tilde{\Gamma}} = 0$, this expression
reduces to the usual BCS formula, 
$T_c = \Lambda \exp ( - 1 / \tilde{\Gamma}_0 )$,
and $\tilde{\Gamma}_c = 0$.
The disorder influences the scaling of the fermion fields,
$d_{\Psi}$, which, in turn, modify $d_{\tilde{\Gamma}}$:
\begin{equation}
d_{\tilde{\Gamma}} = 4 d_{\Psi} - 3 = 1 - \frac{2 \Delta}{\pi}
\end{equation}
where $\Delta $ is given 
in  eq.\{\ref{vortex}\} or eq.\{\ref{disloc}\}. 
The critical temperature depends
exponentially on the disorder. The expression in eq.\{\ref{Tc}\} is
only valid if $T_c \ll v_F / d$, where $d$ is the typical distance
above which eq.\{\ref{vortex}\} or eq.\{\ref{disloc}\} hold.

We can make a simple estimate of the role of disorder by assuming that,
for certain average separation between defects, $l_0$, 
$d_{\tilde{\Gamma}} = 0$, and the value of the critical temperature
is $T_c^{max}$. Then, if the disorder is reduced, we expand
on $d_{\tilde{g}} \ll 1$, and we obtain:
\begin{eqnarray}
T_c^0 &\approx  &\Lambda e^{- \frac{1}{\tilde{\Gamma}_0}} 
e^{- \frac{d_{\tilde{\Gamma}}}{2 \tilde{\Gamma}_0^2}}
\nonumber \\ &\sim &T_c^{max} e^{- k \frac{l - l_0}{l_0}}
\end{eqnarray}
where $k \propto \tilde{\Gamma}_0^{-2}$
is a numerical constant and $l$ is the average distance between defects.
The superscript 0 stands for the fact that frustration effects
in the superconducting phase
are not considered (see below).
Finally, we can get a rough estimate for $l_0$ by considering that
a sufficiently large concentration of defects leads to pair
breaking and reduces $T_c$ in an anisotropic superconductor.
The reduction of $T_c$ is given, approximately, by\cite{G98}:
\begin{equation}
T_c \approx T_c^0 \left( 1 -  c \frac{\xi_0^2}{l_0^2} \right)
\end{equation}
where $\xi_0 = v_F / T_c^0$ is the coherence length of the
superconductor, and $c$ is a constant of order unity.
Hence, the optimal concentration of
defects will be in the range $l_0 \sim \xi_0$. Assuming that
$T_c^{max} \sim 300$K, this estimate gives for the mean
distance between defects $l_0 \sim 30 - 100$\AA.
\subsection{Origin of disorder in graphene sheets.}
It is known that electronic properties of graphite, like the resistivity,
are sample dependent\cite{Eetal98}, and localization effects due to
disorder have been observed\cite{Betal90}. As discussed in section IV,
the effect of topological disorder depends on whether the graphene 
sheets present a finite density of disclinations, leading to
corrugated and warped surfaces, or the main source of disorder
is due to dislocations. 
Aggragations of graphite nanoparticles of polyhedron
shapes, whose curvature
is not completely characterized are discussed 
in\cite{Aetal98}. Warped layers, with curved regions which
are reminiscent of the spherical fullerenes have been observed\cite{Aetal99}.
These structures seem similar to proposed models of negatively curved graphene
layers\cite{Tetal92}. Theoretically, these compounds (schwarzites) are
supposed to be very stable, and contain a macroscopic fraction of
heptagonal rings. A material with these characteristics is probably
best described by a random distribution of disclinations, with
mean separation equal to a few lattice spacings. Calculations of
the electronic density of states of the model proposed in\cite{Tetal92}
show that it loses the semimetallic properties of graphite, in
agreement with the discussion here\cite{HC94}. Compounds with these
characteristics can be good candidates for intrinsic p-wave superconductivity.

From the difference between eqs.\{\ref{vortex}\} and \{\ref{disloc}\}, it is
clear that a non corrugated graphene sheet has a much lower density
of states than a significantly warped one, 
and a reduced tendency towards electronic instabilities. In highly disordered
graphite, however, it is possible that regions with different degrees
of corrugation coexist giving rise to the behavior reported
in\cite{Ketal00,Ketal00b}.

\section{Acknowledgements.}
We are thankful to M. P. L\'opez-Sancho, F. Batall\'an
and P. Esquinazi for helpful
discussions. Financial support from MEC (Spain) through grant
PB96/0875 and CAM (Madrid) through grant 07/0045/98 are gratefully
acknowledged.


\begin{references}
\bibitem{Ketal00}
Y. Kopelevich, P. Esquinazi, J. H. S. Torres and S. Moehlecke,
J. Low Temp. Phys. {\bf 119}, 691 (2000).
\bibitem{Ketal00b}
H. Kempa, Y. Kopelevich, F. Mrowka, A. Setzer, 
J. H. S. Torres, R. H\"ohne and P. Esquinazi,
preprint (cond-mat/0005439).
\bibitem{GGV94}
J. Gonz\'alez, F. Guinea and M. A. H. Vozmediano, 
Mod. Phys. Lett. {\bf B7}, 1593 (1994), {\it ibid},
Nucl. Phys. B
{\bf 424}, 595 (1994), {\it ibid}, Journ. Low. Temp. Phys. {\bf 99},
287 (1995).
\bibitem{S94}
R. Shankar, Rev. Mod. Phys. {\bf 66} (1994) 129. 
\bibitem{P92}
J. Polchinski, in
{\bf Proceedings of the 1992 TASI in Elementary Particle Physics},
J. Harvey and J. Polchinski eds. (World Scientific, Singapore, 1992).
\bibitem{GGV96}
J. Gonz\'alez, F. Guinea and M. A. H. Vozmediano, Phys. Rev. Lett. {\bf
77}, 3589 (1996).
\bibitem{Yetal96}
S. Yu, J. Cao, C. C. Miller, D. A. Mantell, R. J. D. Miller
and Y. Gao, Phys. Rev. Lett. {\bf 76}, 483 (1996).
\bibitem{GGV99}
J. Gonz\'alez, F. Guinea and M. A. H. Vozmediano, Phys. Rev. B,
{\bf 59} R2474 (1999).
\bibitem{GGV00}
J. Gonz\'alez, F. Guinea and M. A. H. Vozmediano, Phys. Rev. Lett.
{\bf 84}, 4930 (2000).
\bibitem{GGV92}
J. Gonz\'alez, F. Guinea and M. A. H. Vozmediano,
Phys. Rev. Lett. {\bf 69}, 172 (1992),
Nucl. Phys. {\bf B406}, 771 (1993).
\bibitem{G98}
F. Guinea, Phys. Rev. B {\bf 58}, 6622 (1998).
\bibitem{Letal94}
A. W. W. Ludwig, M. P. A. Fisher, R. Shankar and G. Grinstein,
Phys. Rev. B {\bf 50}, 7526 (1994).
\bibitem{E83}
E. N. Economou, {\bf Green's Functions in Quantum Physics},
Springer, Berlin (1983).
\bibitem{NTW94}
A. A. Nersesyan, A. M. Tsvelik and F. Wenger, Phys. Rev. Lett.
{\bf 72}, 2628 (1994), {\it ibid}, Nucl. Phys. B {\bf 438}, 561 (1995). 
\bibitem{CMW96}
C. de C. Chamon, C. Mudry and X. G. Wen, Phys. Rev. Lett. {\bf 77},
4194 (1996), {\it ibid}, Nucl. Phys. B {\bf 466}, 383 (1996).
\bibitem{MH97}
Y. Morita and Y. Hatsugai, Phys. Rev. Lett. {\bf 79}, 3728 (1997).
\bibitem{Aetal00}
W. A. Atkinson, P. J. Hirschfeld, A. H. MacDonald and K. Ziegler,
preprint (cond-mat/0005487).
\bibitem{PL98}
C. P\'epin and P. A. Lee, Phys. Rev. Lett. {\bf 81}, 2779 (1998).
\bibitem{SF99}
T. Senthil and M. P. A. Fisher, Phys. Rev. B {\bf 60}, 6893 (1999).
\bibitem{W90}
X.-G. Wen, Phys. Rev. B {\bf 42}, 6623 (1990).
\bibitem{KF92}
C. L. Kane and M. P. A. Fisher, 
Phys. Rev. B {\bf 46}, 15233 (1992).
\bibitem{Z78}
A. Zunger, Phys. Rev. B {\bf 17}, 676 (1978).
\bibitem{ST92}
S. Sorella and E. Tosatti,  Europhys. Lett. {\bf 19} 699 (1992).
\bibitem{TH92}
A. L. Tchougreeff and R. Hoffmann, J. Phys. Chem. {\bf 96},
8993 (1992).
\bibitem{WL96}
F. R. Wagner and M.-B. Lepetit, J. Phys. Chem. {\bf 100}, 11050 (1996).
\bibitem{LCM00}
M. P. L\'opez-Sancho, L. Chico and M. C. Mu\~noz, to be published.
\bibitem{Hetal65}
N. B. Hannay, T. H. Geballe, B. T. Mathias, K. Andres, P. Schmidt and
D. MacNair, Phys. Rev. Lett. {\bf 14}, 255 (1965).
\bibitem{KT80}
Y. Koike and S. Tanuma, J. Phys. Chem Solids, {\bf 41}, 1111 (1980).
\bibitem{JDC91}
R. A. Jishi, M. S. Dresselhaus and A. Chaiken, Phys. Rev. B
{\bf 44}, 10248 (1991).
R. A. Jishi and M. S. Dresselhaus, Phys. Rev. B {\bf 45}, 12465 (1992).
\bibitem{Eetal98}
L. Edman, B. Sundqvist, E. McRae and E. Litvin-Staszewska, Phys. Rev. B
{\bf 57}, 6227 (1998).
\bibitem{Betal90}
V. Bayot, L. Piraux, J.-P. Michenaud, J.-P. Issi, M. Lelaurain and
A. Moore, Phys. Rev. B {\bf 41}, 11770 (1990).
\bibitem{Aetal98}
O. E. Andersson, B. L. Prasad, H. Sato, T. Enoki, Y. Hishiyama,
Y. Kaburagi, M. Yoshikawa and S. Bandow, Phys. Rev. B {\bf 58},
16387 (1998).
\bibitem{Aetal99}
I. Alexandrou, H.-J. Scheibe, C. J. Kiely, A. J. Papworth,
G. A. Amaratunga and B. Schultrich, preprint (cond-mat/9905130).
\bibitem{Tetal92}
S. J. Townsend, T. J. Lenosky, D. A. Muller, C. S. Nichols
and V. Elser, Phys. Rev. Lett. {\bf 69}, 921 (1992).
\bibitem{HC94}
M.-Z. Huang and W. Y. Ching, Phys. Rev. B {\bf 49}, 4987 (1994).
\end{references}
\end{document}